\documentclass[preprint,aps,prl,onecolumn,superscriptaddress]{revtex4}  % for review and submission
\usepackage{graphicx}  % needed for figures
\usepackage{dcolumn}   % needed for some tables
\usepackage{bm}        % for math
\usepackage{amssymb}   % for math
\usepackage{amssymb,amsmath}   % for math
\usepackage{txfonts}
\usepackage{upgreek}
\usepackage{CJK,graphicx}
\usepackage{color}

\begin{document}

% the following line is for submission, including submission to the arXiv!!
%\hspace{5.2in} \mbox{Fermilab-Pub-04/xxx-E}

\title{Lattice-site-specific spin-dynamics in double perovskite Sr$_2$CoOsO$_6$}
\author{Binghai Yan}
\affiliation{Max-Planck-Institut f\"{u}r Chemische Physik fester Stoffe, 01187 Dresden, Germany}
\affiliation{Max-Planck-Institut f\"{u}r Physik komplexer Systeme, 01187, Dresden, Germany}
\author{Avijit Kumar Paul}
\author{Sudipta Kanungo}
\affiliation{Max-Planck-Institut f\"{u}r Chemische Physik fester Stoffe, 01187 Dresden, Germany}
\author{Manfred Reehuis}
\author{Andreas Hoser}
\author{Daniel M. T\"{o}bbens}
\affiliation{Helmholtz-Zentrum f\"{u}r Materialien und Energie, 14109 Berlin, Germany}
\author{Walter Schnelle}
\affiliation{Max-Planck-Institut f\"{u}r Chemische Physik fester Stoffe, 01187 Dresden, Germany}
\author{Robert C. Williams}
\author{Tom Lancaster}
\author{Fan Xiao}
\affiliation{Durham University, Department of Physics, South Road, Durham, DH1 3LE UK}
\author{Johannes S. M\"{o}ller}
\author{Stephen J. Blundell}
\author{William Hayes}
\affiliation{University of Oxford, Department of Physics, Clarendon Laboratory, Parks Road, Oxford, OX1 3PU UK}
\author{Claudia Felser}
\affiliation{Max-Planck-Institut f\"{u}r Chemische Physik fester Stoffe, 01187 Dresden, Germany}
\affiliation{Johannes Gutenberg-Universit\"{a}t, Institut f\"{u}r Anorganische Chemie und Analytische Chemie, 55128 Mainz, Germany}
\author{Martin Jansen}
\email{m.jansen@fkf.mpg.de}
\affiliation{Max-Planck-Institut f\"{u}r Chemische Physik fester Stoffe, 01187 Dresden, Germany}
\affiliation{Max-Planck-Institut f\"{u}r Festk\"{o}rperforschung, 70569 Stuttgart, Germany}

\date{\today}

\begin{abstract}
The magnetic properties and spin-dynamics have been studied for the structurally-ordered double perovskite Sr$_2$CoOsO$_6$. Neutron diffraction, muon-spin relaxation and ac-susceptibility measurements reveal two antiferromagnetic (AFM) phases on cooling from room temperature down to 2~K. In the first AFM phase, with transition temperature $T_{\mathrm{N1}}=108$~K, cobalt (3$d^7$, $S=3/2$) and osmium (5$d^2$, $S=1$) moments  fluctuate dynamically, while their \textit{average} effective moments undergo long-range order. In the second AFM phase below $T_{\mathrm{N2}}=67$~K, cobalt moments first become frozen and induce a noncollinear spin-canted AFM state, while dynamically fluctuating osmium moments are later frozen into a randomly canted state at $T\approx 5$~K. \textit{Ab initio} calculations indicate that the effective exchange coupling between cobalt and osmium sites is rather weak, so that cobalt and osmium sublattices exhibit different ground states and spin-dynamics, making Sr$_2$CoOsO$_6$ distinct from previously reported double-perovskite compounds.
\end{abstract}

\maketitle

%Introduction
The structural and electronic properties of solids are sensitively affected by the subtle balance between microscopic exchange interactions and other competing factors. When magnetic exchange interactions cannot be simultaneously satisfied on a lattice (a situation called {\it frustration}), long-range order can be suppressed and new cooperative phenomena may emerge, such as spin-liquid~\cite{Anderson:1973eo,Balents:2010ds}, spin-glass~\cite{Mydosh1993,Binder:1986vt} and spin-ice~\cite{Bramwell:2001ex,Anonymous:4JcjUQiO} states. 
Some special lattice geometries with antiferromagnetic coupling host an intrinsic frustration effect ~\cite{Ramirez:1994wx,Schiffer1996,Moessner:2001cf}. The double-perovskite compound $A_2BB^\prime$O$_6$ is a simple model system showing geometrical frustration~\cite{Karunadasa:2003bg}.  Here $A$ is usually an alkaline-earth or rare-earth element, and $B$ and $B^\prime$ are transition-metal elements. The system consists of two interpenetrating $fcc$ sublattices, $B$ and $B^\prime$, both of which are composed of an edge-shared network of  tetrahedra, typical units of frustration. However, the inter-sublattice coupling is also significant due to the sublattice interpenetration. Thus, the interplay between intra-lattice frustration and the inter-lattice interaction leads to rich magnetic behavior including ferromagnetism (FM), antiferromagnetism (AFM), multiferroicity, spin-liquids and spin-glass states~\cite[and references therein]{Serrate:2005jw,Paul:2013dy,Tokura:1998ke,Battle:1989bp}. Since considerable spin-orbit coupling of $5d$ states promises exotic phases in double-perovskites~\cite{Chen:2010ec,Chen:2011wh,Meetei:2013fd}, compounds with $B^\prime = $ Re and Os have attracted much research interest. For example, Sr$_2$CrOsO$_6$~\cite{Krockenberger:2007vy} and Sr$_2$FeOsO$_6$~\cite{Paul:2013et,Paul:2013dy} exhibit the 5$d^3$ configuration while Ba$_2$CaOsO$_6$~\cite{Yamamura:2006ux}, Sr$_2$CrReO$_6$ and Ba$_2$YReO$_6$~\cite{Aharen:2010kw} show 5$d^2$.

In this Letter, we report exotic lattice-site-specific spin-dynamics in the double-perovskite Sr$_2$CoOsO$_6$, in which Co (3$d^7$, $S=3/2$) and Os (5$d^2$, $S=1$)  moments show different spin dynamics and freezing processes and eventually condense into different ground states. Combining neutron diffraction with muon spin relaxation ($\mu^{+}$SR) and ac-susceptibility (ac-$\chi$) measurements, we find that two antiferromagnetic phases evolve from the paramagnetic phase on cooling from room temperature down to 2~K. In the first AFM phase, long-range order is observed by neutron powder diffraction, where averaged moments from dynamical (partially frozen) Co and Os spins orient along the [110] direction. In the second AFM phase, Co moments first become totally frozen and induce a new spin-canted noncollinear  AFM state while Os moments continue to exhibit dynamics. Upon cooling further down to $T\approx 5$~K, Os moments become frozen into a randomly canted state, even though their averaged moments still preserve AFM order. Our \textit{ab initio} calculations indicate that the $fcc$-like Os sublattice shows strong geometrical frustration and couples weakly to the Co sublattice, which accounts for its glassy, randomly-frozen spins. The details of experimental and calculation methods can be found in the supplementary information (SI).

\begin{figure*}
  \centering
  \includegraphics [width=16cm] {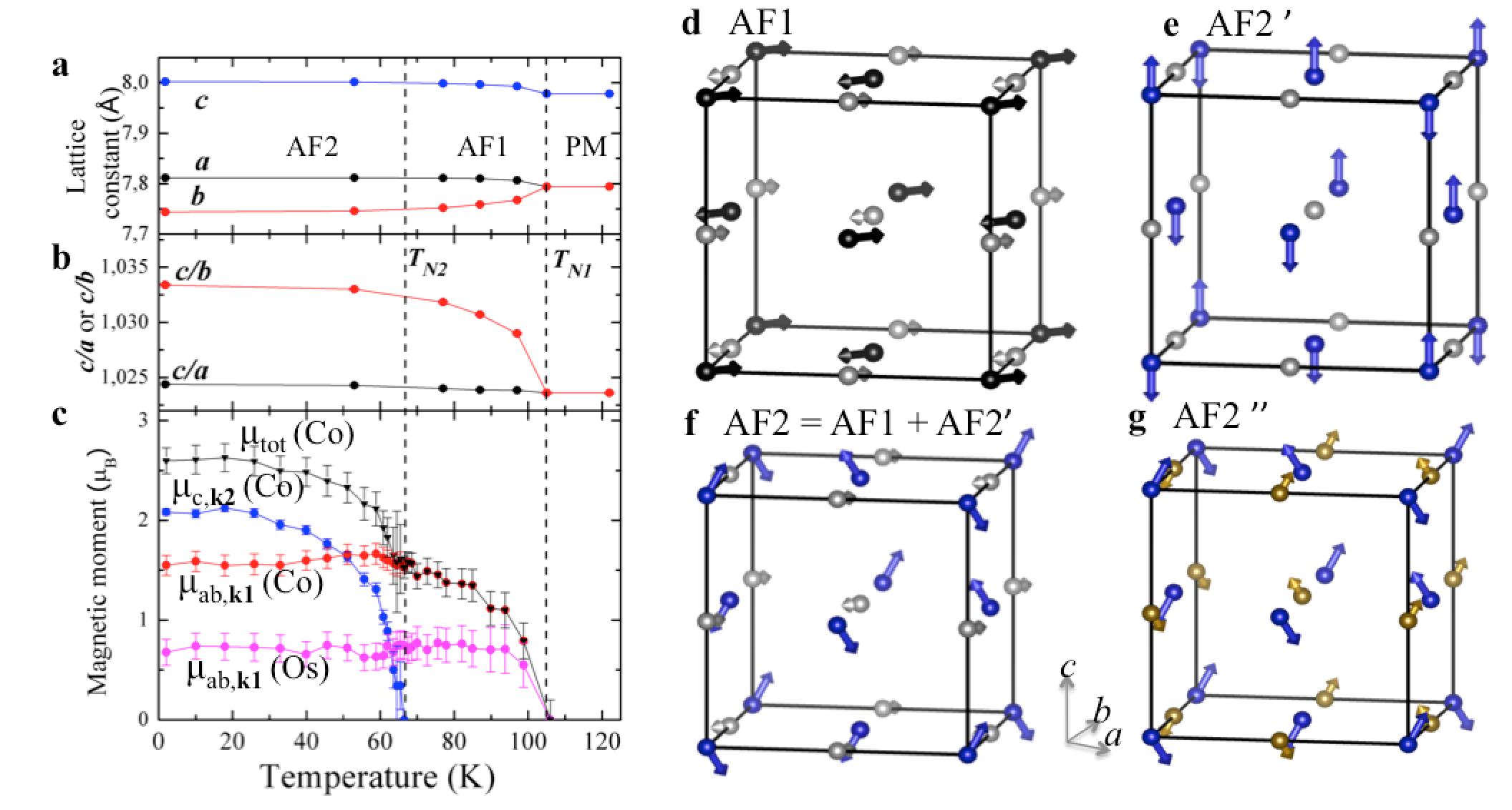}\\
  \caption {(Color online) Results of the neutron powder diffraction study. (a) $-$ (b) Temperature dependence of lattice parameters $a$, $b$ and $c$, as well as their ratios $c/a$ and $c/b$. (c) Temperature dependence of the magnetic moments of Co and Os atoms.  (d) $-$ (g) Magnetic structure of the AF1 (refined with $\boldsymbol{k_1}$), AF2$^\prime$ (refined with $\boldsymbol{k_2}$), AF2 and AF2$^{\prime \prime}$ phases (see the text), where Os spins are canted randomly in AF2$^{\prime \prime}$. Black and blue spheres (arrows) represent Co atoms (corresponding moments) with dynamical and frozen spins, respectively. Similarly, gray and yellow spheres (arrows) represent Os atoms (corresponding moments) with dynamical and frozen spins, respectively. }
\end{figure*}

% Neutron powder diffraction

The synthesized Sr$_2$CoOsO$_6$ exhibits two phase transitions based on previous susceptibility and heat-capacity measurements~\cite{Paul:2013ui}, one from paramagnetic to an AFM phase (labelled as AF1) at the N$\mathrm{\acute{e}}$el temperature $T_{\mathrm{N}1}$ = 108~K, and the other from the AF1 phase to a second AFM phase (labelled AF2) at $T_{\mathrm{N}2} = 67$~K. To investigate the exact spin configurations, magnetic and crystal lattice structures were investigated by neutron powder diffraction measurements made between 2 and 122~K (Fig.~1), well below and above the structural and magnetic phase transition temperature of 108~K. For the paramagnetic phase above $T_{N1}$, a tetragonal structure is observed as a fully ordered lattice with space group $I4/m$ (No.\ 87). In the AF1 phase, the crystal structure exhibits monoclinic distortions in the $ab$-plane leading to the lower symmetric space group $I2/m$ (No.\ 12) ~\cite{Paul:2013ui}. However, in the following, we shall use a pseudo-cubic unit cell of the monoclinic space group $B2/n$ (No.\ 15)~\cite{Howard2003}, instead of the pseudo-tetragonal one of $I2/m$, in order to specify the magnetic lattice conveniently and concisely.

When the temperature decreases, the lattice distorts further, e.g.\ indicated by the increasing $c/b$ ratio [Figs.~1(a) and 1(b)]. In the neutron powder diffraction pattern, the strongest magnetic reflection is observed at 2$\theta$ = 13.5$^{\circ}$ (see details in SI). Using the monoclinic unit cell, this reflection is indexed as (0,1,0)$_M$, which indicates the presence of ferro- or ferrimagnetic planes perpendicular to the $b$ axis. All the observed magnetic reflections obey the rule ($hkl$)$_M$ = ($hkl$)$_N \pm \boldsymbol{k}$ with  $\boldsymbol{k}=\boldsymbol{k_1}=$ (0,1,0). This is an $A$-type of AFM order as illustrated in Fig. 1d. The refinement of the magnetic structure shows that the moments of Co and Os atoms are aligned within the $ab$ plane reaching at 70 K moment values of $\mu_{ab,\boldsymbol{k_1}}$(Co) = 1.6(1)~$\mu_{\mathrm{B}}$ and $\mu_{ab,\boldsymbol{k_1}}$(Os) = 0.7(1)~$\mu_{\mathrm{B}}$. 
 Very similar values to those found from our refinements were reported recently by Morrow {\it et al.}~\cite{Morrow:2013ka}. However, they proposed that only the Os sites are ordered in the AF1 phase, yielding a moment $\mu_{ab} = 1.6 ~\mu_{\mathrm{B}}$, which is assigned to the cobalt atoms in our work. Their representation analysis shows that the spin structure they propose is compatible with the irreducible representation $\Gamma_1$ of ours, which contains two basis vectors in the $ab$-plane.
Further, the moments of both metal ions form an angle to the $a$ axis of 43(6)$^{\circ}$, where they are aligned almost parallel to the [110] direction, demonstrating considerable magnetic anisotropy. 
% the second AFM phase
For $T< 67$~K, the $\boldsymbol{k_1}$ type of reflection and corresponding Co and Os moments remain almost unchanged, while a second set of magnetic reflections appears with a propagation vector $\boldsymbol{k}=\boldsymbol{k_2}$= ($\frac{1}{2},\frac{1}{2},\frac{1}{2}$), where the strongest reflection observed at 2$\theta$ = 11.5$^{\circ}$ is indexed as ($\frac{1}{2},\frac{1}{2},\frac{1}{2}$)$_M$. This suggests the presence of a second magnetic phase AF2, in which magnetic moments are treated as a vector sum between those of $\boldsymbol{k_1}$  and $\boldsymbol{k_2}$ types of reflections. 
We use AF2$^\prime$ to label the pseudo-magnetic phase refined with $\boldsymbol{k_2}$ and hence, AF2 can be regarded as a superposition between AF1 and AF2$^\prime$, i.e.\ AF2 = AF1 + AF2$^\prime$. AF2$^\prime$ gives a magnetic unit cell with the dimensions $2a \times 2b \times 2c$. The refinement of the magnetic structure for AF2$^\prime$ in Fig.~1(e) shows that Co moments are aligned preferably parallel to the $c$ axis, 
while the refined moment $\mu_{ab,\boldsymbol{k_2}}$(Os) = 0.1(2)~$\mu_{\mathrm{B}}$ shows that there are no ordered moments on the Os sites.
We note that it is suggested
 in Ref.~\onlinecite{Morrow:2013ka}  that the cobalt moments align within the $ac$ plane. However, an interpretation that is more consistent
with our data from other techniques (discussed below) is obtained if the moments in the AF1 phase align within the $ab$ plane and an additional ordering in AF2$^\prime$ occurs along $c$.
At 2~K the $c$ component of the Co moment reaches a value of $\mu_{c,\boldsymbol{k_2}}$(Co) = 2.1(1)~$\mu_{\mathrm{B}}$, resulting finally in a total magnetic moment $\mu_{\mathrm{tot}}$(Co) = 2.6(1)~$\mu_{\mathrm{B}}$ via a vector sum over its $ab$ and $c$ components. 
However, the moment of the osmium atoms $\mu_{\mathrm{tot}}$(Os) = 0.7(1)~$\mu_{\mathrm{B}}$ remains almost unchanged between 2 and 70~K, since Os moments are found to be zero in AF2$^\prime$. 
It should be noted that we use two propagation vectors, $\boldsymbol{k_1}$ and $\boldsymbol{k_2}$, to refine the AF2 phase, so that one can easily track the evolution from the AF1 to AF2 phases. In principle, using the magnetic unit cell with the dimensions $2a \times 2b \times 2c$, these two magnetic components, $\mu_{ab,\boldsymbol{k_1}}$ and $\mu_{c,\boldsymbol{k_2}}$, could be refined simultaneously as a single magnetic phase with $\boldsymbol{k}$ = 0. In this phase, along the three pseudo-cubic axes one finds for the $ab$ and $c$ components spin sequences of $+ + + + \ldots$ and $+ - + - \ldots$, while along the [110] direction one finds $+ - + - \ldots$ and $+ + + + \ldots$. As shown in Fig.~1(f), the AF2 phase exhibits a noncollinear magnetic structure, in which Co moments are canted in the $c$ direction while Os moments stay in the $ab$ plane.
    
In our calculations, magnetic moments for Co and Os are found to be 2.7~$\mu_{\mathrm{B}}$ and 1.6~$\mu_{\mathrm{B}}$ (details can be found in the SI), respectively. The saturated $\mu_{\mathrm{tot}}$(Co) in AF2 almost reaches the theoretical value, while $\mu_{\mathrm{tot}}$(Co) in AF1 and $\mu_{\mathrm{tot}}$(Os) in both AF1 and AF2 are much smaller than corresponding calculated moments. Since neutron diffraction measures ordered spins over considerable scattering events, only the \textit{averaged} moments can be observed if spins are dynamically fluctuating and at least partially ordered, which should be smaller than the full values. Therefore, we propose the following scenario to interpret the observed AF1 and AF2 magnetic structures. In the AF1 phase with $T_{\mathrm{N}1}>T>T_{\mathrm{N}2}$ both Co and Os spins are dynamical fluctuating with the in-plane moments $\mu_{ab,\boldsymbol{k_1}}$ partially ordered. When cooling down into the AF2 phase, the $c$ components of Co spins start to become frozen for $T<T_{\mathrm{N}2}$ and become totally static for $T < 20$~K, forming the canted noncollinear configuration. In contrast, Os spins become  frozen into a randomly canted state [illustrated as AF2$^{\prime\prime}$ in Fig.~1(g)] for $T<5$~K, so that no long-range order is observed in AF2$^\prime$ at temperature down to 2~K. In order to validate this picture of spin-dynamics, we further performed ac-$\chi$ and $\mu^+$SR measurements.  It should be pointed out that the glass-like
behavior that we observe at low temperature, described in more detail below,
lends weight to our interpretation of the magnetic configurations based
on the neutron data. This contrasts with the spin configurations
presented in Ref. ~\onlinecite{Morrow:2013ka}  where the proposed AFM order below 70 K appears incompatible with
the glassiness we observe.

% ac & muon-SR

The proposed low-temperature glassy behavior of Os spins is reminiscent of a spin-glass~\cite{Binder:1986vt}.  
We can get an insight into the spin-dynamics at lower temperature ($T \ll 67$~K) through the ac-susceptibility.
The ac-$\chi(T,\omega)$ measurement can monitor the freezing process by observing the $\tau - T_{\mathrm{f}}$ or 
$\omega-T_{\mathrm{f}}$ relation~\cite{Binder:1986vt}. As shown in Fig.~2(a), a clear cusp around 6~K and the shift in the peak position of ac-$\chi^\prime (T,\omega)$ [the real part of ac-$\chi (T,\omega)$] from 6~K to 6.8~K with increasing $\omega$, are observed. %
Then one can obtain $\tau$ through frequencies and the corresponding freezing temperature $T_{\mathrm{f}}$ through the peak positions of ac-$\chi^{\prime}(T,\omega)$. For a standard critical slowing down~\cite{Hohenberg:1977fq}, the spin-relaxation time follows a power-law divergence of the form $\tau=\tau_0[T_{\mathrm{f}}/(T_{\mathrm{f}}-T_{\mathrm{f}0})]^{z\nu}$ , where $\tau_0$ is the characteristic time scale of the spin-relaxation and $z\nu$ are critical exponents. The best fit to the above equation, as shown in the right inset of Fig.~2(a), gives $T_{\mathrm{f}0}$ = 5~K, $z\nu$ = 6.1 and $\tau_0 = 4\times 10^{-6}$~s. We note that $\tau_0$ values of $10^{-12}$~s are often observed in a typical spin-glass system ~\cite{Mydosh1993,Binder:1986vt}. The high value of $\tau_0$ indicates relatively slower spin dynamics in our case. The time-dependent ac-$\chi$ confirms that Os spins get frozen into a random canted states where the averaged moments still preserve the AF1 order, in agreement with neutron diffraction results.

\begin{figure}
\centering
\includegraphics [width=16 cm] {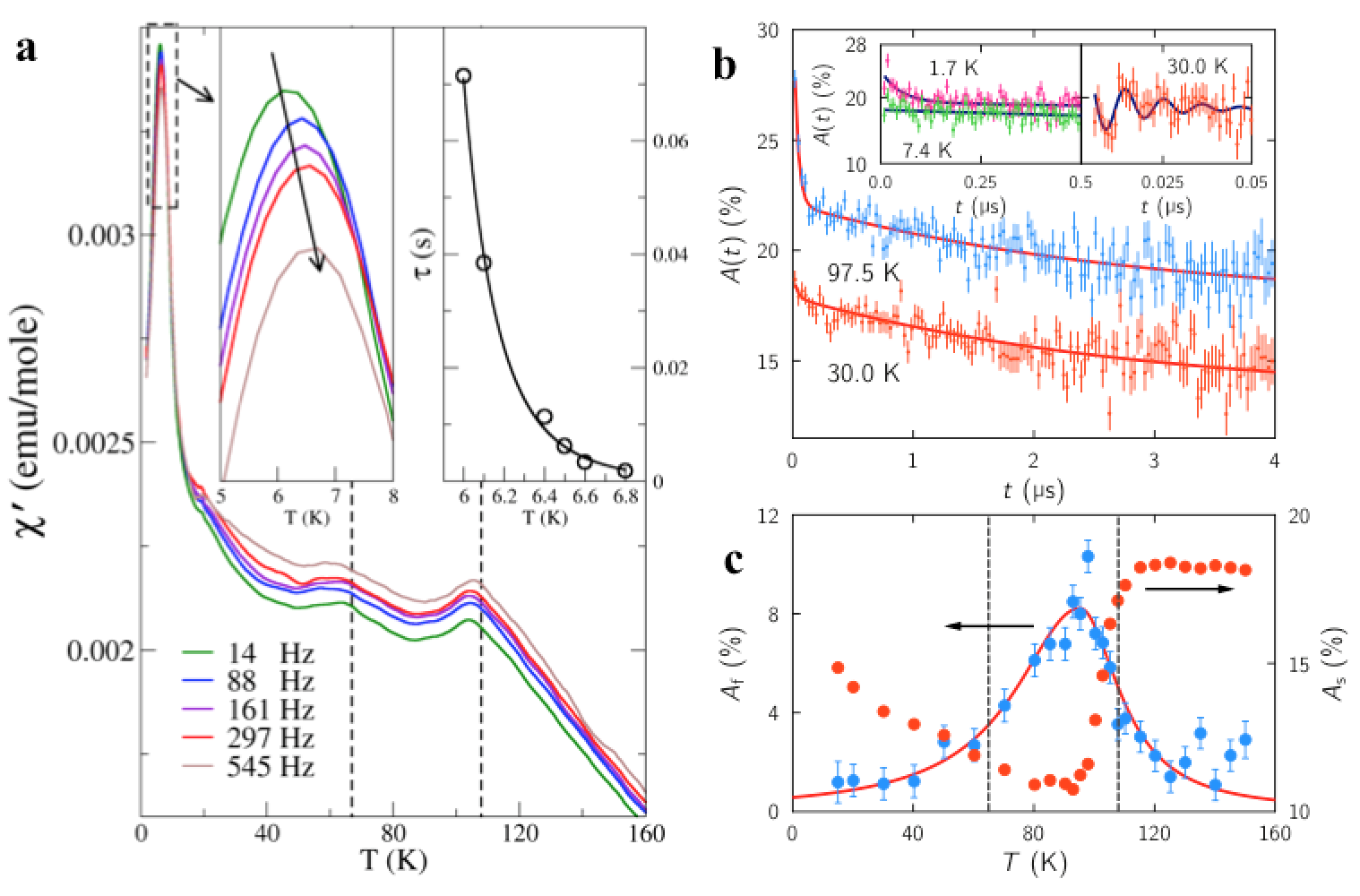}\\
\caption{ (Color online) The ac-susceptibility and muon-spin relaxation measurements. (a) The real part of ac-susceptibility (ac-$\chi^\prime(T, \omega)$). Two transition temperatures are indicated by vertical dashed lines. The left inset shows the peak shift with respect to different frequencies $\omega$. The right inset shows the $T_{\mathrm{f}}$ dependence on $\tau=1/\omega$, where the solid line represents the fitting curve (see main text). (b) Muon-spin relaxation spectra measured for $T<T_{\mathrm{\mathrm{N}1}}$. The
spectrum measured at $T=30$~K is offset by 6\% for clarity. 
In the region $65 \leq T \leq 110$~K a fast relaxing component
is observed at early times which vanishes as the temperature is lowered. 
{\it Left inset:} Cooling below 2~K results in a sharp increase in
non-relaxing amplitude and additional relaxation becoming resolvable. 
{\it Right inset:} Below 65~K oscillations are observed consistent with a
uniform magnetic field distribution resulting from the ordered Co sublattice
only. (c) The amplitude of the fast relaxing component is seen to
 peak around 100~K and decreases as temperature is lowered further, 
while the amplitude of a non-relaxing component increases on cooling
below this point. \label{muonfig}}
\end{figure}

Muon spin relaxation ($\mu^{+}$SR) measurements also support the picture from neutron diffraction and ac-$\chi$. For $T\gg 110$~K only slow relaxation is observed, typical of a paramagnet. On cooling below $T\approx 110$~K 
a fast-relaxing component is resolved (see Fig.~\ref{muonfig}b) with a sizeable relaxation rate, resulting from the muon ensemble experiencing a broad distribution
of large, slowly fluctuating magnetic fields. Its development
coincides with a loss of asymmetry in the amplitude of a non-relaxing
(or, more likely, very slowly relaxing)
component $A_{\mathrm{s}}$.  (However, the total observed asymmetry is reduced
compared to its value in the paramagnetic regime.)
The data were fitted to a relaxation function $A(t) = A_{\mathrm{s}} + A_{\mathrm{f}}{\rm e}^{-\Lambda t} +
A_{\mathrm{m}}{\rm e}^{-\lambda t}$, with fixed parameters
$A_{\mathrm{m}}=4.5$\%,  $\lambda=0.35$~MHz and $\Lambda=30$~MHz, 
determined from an initial fit of the data where all parameters were allowed
to vary. The component with amplitude $A_{\mathrm{m}}$ therefore represents a
temperature independent contribution to the signal. 
The amplitude $A_{\mathrm{f}}$ peaks
below 100~K (Fig.~\ref{muonfig}c) and then smoothly decreases as temperature is lowered further. The disappearance of this fast-relaxing component is
accompanied by the smooth increase in $A_{\mathrm{s}}$.
Below a temperature $T\approx 65$~K we also
begin to resolve spontaneous oscillations in the $\mu^{+}$SR spectra
at very early times (right inset, Fig.~\ref{muonfig}b), 
consistent with the existence of quasistatic field distribution at the muon sites,
with a narrow width, involving smaller and more uniform fields than
that giving rise to the high temperature behavior. Fitting these to a
cosinusoidal function allows us to estimate the oscillation frequency
as $\nu_{\mu} \approx 90$~MHz. 

It is likely that the muon is sensitive to the 
field distribution resulting from both Co and Os moments, although the
time-scale of the muon measurement differs from that of 
neutron diffraction.  Compared to muons, whose time-scale is determined by the gyromagnetic ratio $\gamma_{\mu} = 2 \pi \times 135.5$~MHz T$^{-1}$, 
neutrons effectively take a `snap-shot' of the spin
distribution (averaged over many scattering events), so will see magnetic order in cases where the muon response reflects slow fluctuations of the moments on the nanosecond or microsecond timescale. As a result, the fast relaxation with amplitude $A_{\mathrm{f}}$ most probably  reflects the slowing down and freezing of the Co moments. Although
these moments undergo a transition to become (partially) ordered on the neutron timescale below 108~K, 
slow fluctuations appear to prevent a discontinuous response of the muon probe
at this temperature. Instead we see an increase in $A_{\mathrm{f}}$ below 110~K, which
probably corresponds to partial order of the Co moments developing on cooling
until around 100~K. Below that temperature, we have the partial order of the Co ions on the muon time
scale,  coexisting with the fluctuations along
the $z$-direction which themselves slow down on cooling, freezing out
relatively smoothly and causing a decrease in $A_{\mathrm{f}}$. 
In the fast fluctuation limit we expect the muon relaxation rate to be
given by $\lambda = \gamma_{\mu}^{2}\langle B^{2}\rangle
\tau$. 
Given an oscillation frequency ($\nu_{\mu}= \gamma_{\mu}B/2 \pi)$ at low temperature of roughly
$90$~MHz, we may estimate $B \approx 0.6$~T. Using
$\Lambda = 30$~MHz in the region  $65 \lesssim T \lesssim 110$~K we
may then crudely estimate a characteristic timescale for the Co moment fluctuations as $\tau \approx 0.01$~ns. 

Below 65~K the more uniform distribution of
internal magnetic fields that arise from the well-ordered
Co-sublattice allows oscillations in the muon spectra to be resolved.
The increase in the component
$A_{\mathrm{s}}$ on cooling below 100~K might then result from the freezing out
of relaxation channels reflecting dynamic fluctuations of the
partially ordered, fluctuating Os spins.  Finally, on cooling below 2~K (left inset, Fig.~\ref{muonfig}b) we observe
additional relaxation and, most notably, a sharp increase in the 
non-relaxing amplitude of the signal. 
In a powder sample we expect 1/3 of muon spins to lie along the direction of any static local fields and, in the absence of dynamics, not to experience any relaxation. The transition from a dynamic to a static regime may then be identified by such an upward shift in  the non-relaxing amplitude of the signal. 
The observed jump in the non-relaxing amplitude and additional relaxation therefore reflects a freezing of the Os moments on the muon time-scale, which also allows further relaxation processes to enter the muon time-window.

%%%%%%% calculations

\begin{figure}
\centering
\includegraphics [width=16 cm] {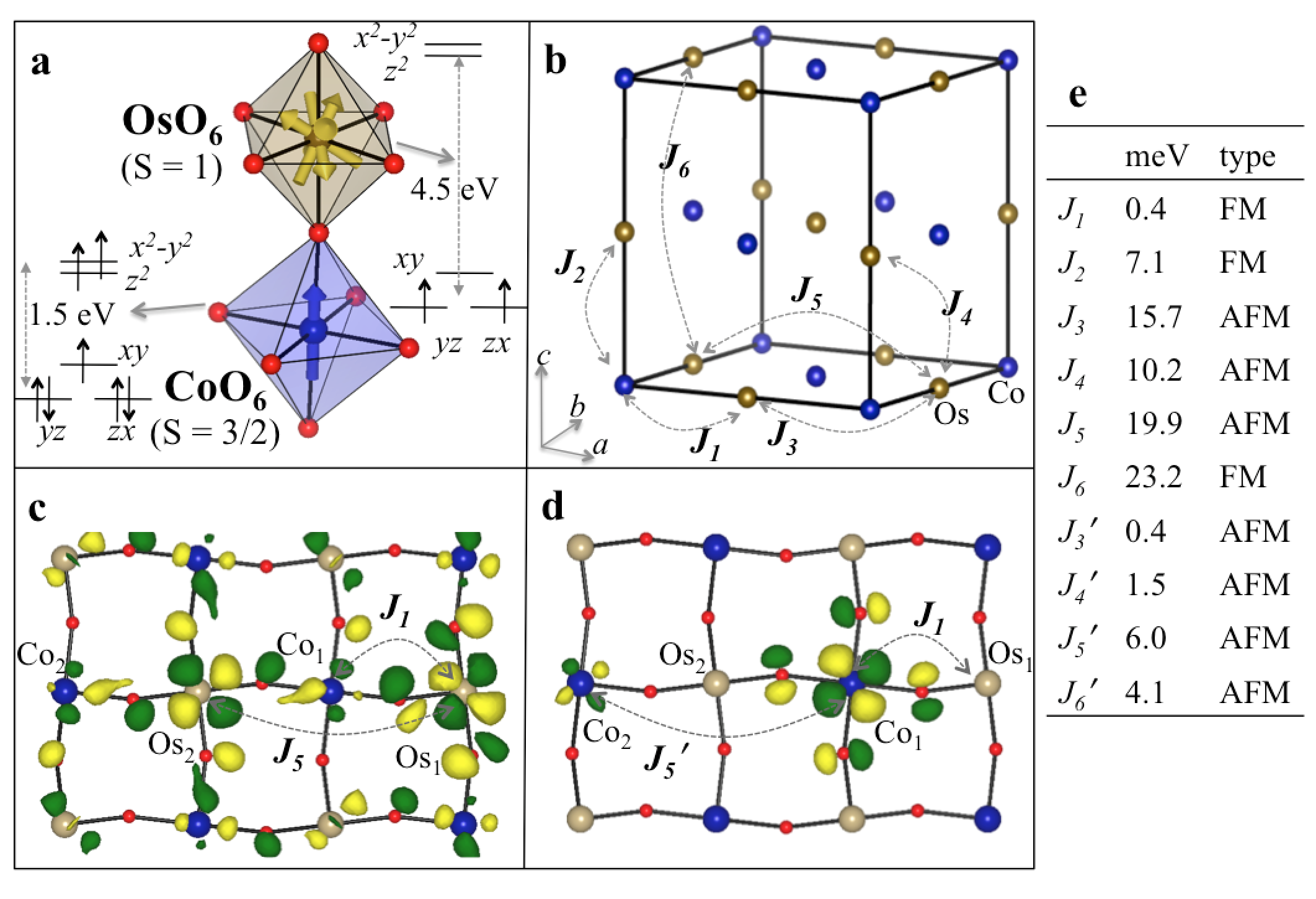}\\
\caption{ (Color online) Electronic and magnetic structures of Sr$_2$CoOsO$_6$ from theoretical calculations.  (a) The local octahedral structures and $d$-orbital diagram. The $t_{2g}–e_g$ splitting and occupations are plotted with Co-3$d^7$ and Os-5$d^2$ configurations. The yellow arrows at the Os site illustrate a random spin orientation, the blue arrow at the Co site represent an ordered state. (b) The exchange coupling pathways in the lattice. Blue and yellow spheres represent Co and Os atoms, respectively. Os and Co forms two interpenetrating $fcc$ sublattices. Sr and O atoms are not shown. Exchanges interactions from $J_3$ to $J_6$ represent  Os$-$Os interactions, while Co$-$Co interactions (from $J_3^\prime$ to $J_6^\prime$ ) are similar to Os$-$Os ones between equivalent lattice sites, which are omitted for the sake of simplicity. The calculated values of these effective exchange interactions are listed in (e). (c)-(d) The Wannier functions of Os-$5d_{xy}$ that centers at Os$_1$ site and Co-3$d_{xy}$ that centers at Co$_1$ site in the $ab$ plane, respectively. Green and yellow surfaces stand for the surfaces with  isovalues +0.15 and -0.15, respectively. Small red spheres represent O atoms.}
\end{figure}

We have shown that cobalt and osmium spins condense into different types of magnetic structure for $T<T_{N2}$, although cobalt and osmium ions form two equivalent $fcc$-like sublattices in the double-perovskite structure. How do these two sublattices become decoupled into different magnetic ground states? To answer this question, we performed \textit{ab initio} calculations to investigate the electronic structure of Sr$_2$CoOsO$_6$. Our calculations reveal Co$^{2+}$ ($3d^7$, $S = 3/2$) and Os$^{6+}$ (5$d^2$, $S = 1$) charge states, which agrees well with experiment~\cite{Paul:2013ui}.  %
We note that the Co-$3d$ and Os-$5d$ states show very little overlap in energy, as illustrated in Fig.~3(a) and the density of states (see Fig. S3 in SI). This implies a weak magnetic exchange coupling between the Co and Os sublattices, which can also be visualized from the Wannier orbitals. The exchange coupling $J$ is approximately proportional to $t^2/U$ in the limit of large Coulomb correlation $U$, where $t$ is the hopping integral between corresponding two sites. Thus, a large orbital overlap may lead to a large hopping value $t$ and hence, a large $J$ value. For instance in Fig. 3c, the Wannier orbital Os-$5d_{xy}$ centering at the Os$_1$ atom, which is projected from \textit{ab initio} wave functions, shows a larger tail at the Os$_2$ site than that at the neighboring Co$_1$ site, indicating that orbital overlap between Os$_1$ and Os$_2$ sites which corresponds to $J_5$ is much stronger than that between Os$_1$ and Co$_1$ sites which corresponds to $J_1$. Similarly Co$_1 -$Co$_2$ $J_5 ^\prime$ is also rather larger than $J_1$, as indicated by Fig. 3d. This is consistent with the inferences above based on the overlap of the density of states. Furthermore, our GGA calculations of the exchange J values (Fig. 3e) indicated in Fig. 3b directly support the above conclusions. Here the Os sublattice is found exhibit much stronger magnetic frustration than the Co sublattice.

In summary, the double-perovskite Sr$_2$CoOsO$_6$ exhibits different magnetic structures as well as spin-dynamics on the Co and Os sublattices, which couple weakly to each other via magnetic exchange interactions. On cooling to low temperature, Co spins become frozen, first into a noncollinear antiferromagnetic state. However, Os spins are frozen into a randomly canted state, which is plausibly due to the dramatic geometrical frustration effect within the Os sublattice, although the averaged $ab$ components of Os moments exhibit an antiferromagnetic order.

\begin{acknowledgments}
We are grateful to R. Moessner, A. K. Nayak, P. Adler and J. Mydosh for fruitful discussion, to A. Amato and H. Luetkens for technical assistance and to the Swiss Muon Source, Paul Scherrer Institut, Switzerland, for provision of muon beamtime. This work is financially supported by the ERC Advanced Grant (291472) and by EPSRC (UK).

\end{acknowledgments}

\end{document}